\documentstyle [12pt,epsfig] {article}
\def\gsim{\buildrel > \over\sim}
\def\lsim{\buildrel < \over\sim}
\def\Pom{I \hspace {-0.14cm} P}
\def\pom{I \hspace {-0.07cm} P}
\begin{document}

\hspace{11cm} {\large {JINR-E2-98-136} }

\vspace{1.0cm}

\begin{center}

{\large { \bf Coherent and Non-Coherent Double Diffractive
Production of $ Q \bar {Q} $ - pairs in Collisions of Heavy
Ions at High Energies.}}

\vspace{1.0cm}

N.M.Agababyan, S.A.Chatrchyan, A.S.Galoyan,
L.L.Jenkovszky {\footnote {\it Bogolyubov Institute for Theoretical Physics
Kiev, Ukraine. }}
A.I.Malakhov, G.L.Melkumov, P.I.Zarubin

\vspace{1.0cm}

{\it Joint Institute for Nuclear Research}

{\it Dubna, Russia}

\vspace{2.0cm}

{\bf Abstract}

\end{center}

The double coherent and non-coherent diffractive production
of heavy quark - antiquark pairs ($Q \bar{Q}$) in heavy ion scattering at high
energies (LHC) is considered. The total and differential cross sections of these
processes with the formation of $c \bar{c}$ and $b \bar{b}$ pairs in $pp$,
$CaCa$ and $PbPb$ collisions are evaluated.

The contribution of the
considered mechanisms is a few per cent of the number of heavy quark -
antiquark pairs obtained in the processes of hard (QCD) scattering, and
it will be taken into account in the registration of $c$, $b$ quarks or, for
instance, in the study of the heavy quarkonia suppression effects in Quark -
Gluon Plasma, in the search for intermediate mass Higgs bosons and so on.

It is shown that the cross section of the coherent scattering process is great
enough. This makes it suitable for studying collective effects in nuclear
interactions at high energies. An example of such effects is given:
large values of the invariant mass of a $Q \bar{Q}$pair,
$M_ {Q \bar {Q}} \gsim 100$GeV, in association with a large rapidity gap
between diffractive jets $\Delta \eta > 5$.

\newpage

{\bf \Large Introduction}

\vspace{0.8cm}

Over the last decade, interest in studying diffractive processes has increased
again. The pronounced features of diffractive (nonperturbative) interactions
were observed in the events registered in the experiments carried out at CERN
\cite{ua8}, on HERA \cite{h1_1}, \cite{zeus_1} and Tevatron \cite{cdf_1},
\cite{d0_1}. First, secondary beams with a large longitudinal momentum
($x_F \gsim 0.9$), i.e. with the ejection of a main energy of initial hadrons
in a narrow phase space volume (diffraction cone), were registered. Second, an
interval in the pseudorapidity space (between these beams), not filled with
secondary hadrons ("rapidity gap"), was observed. ( As shown, the existence of
rapidity gaps (RG) is due to the exchange by a colourless object: a photon,
$W$ -, $Z$ - bosons etc. and, in particular, a Pomeron $\Pom$
\cite{doksh_gap}).  Thus, the energy characteristics of the particles
involved in the process of scattering were such that these processes fell
within the area of QCD applicability. The fraction of such events was: $\sim
6 \div 7 \%$ for $ep$ interactions \cite{h1_1}, \cite{zeus_1} and $\sim 1 \%$
for $p \bar{p}$ interactions \cite{cdf_1}, \cite{d0_1} of the total number of
deep inelastic scattering events. (The theoretical estimates predict a growth
of this value for LHC energies up to $\sim 10 \div 15 \%$, see, for example,
\cite{bjerken_glob}, \cite{land_303}). A detailed analysis of the experimental
data has shown that they are well described by the assumption of Pomeron
exchange.

The hypothesis of Pomerons, first suggested by I.Ya.Pomeranchuk in 1958,
was used to explain the behaviour of the total cross section of hadron - hadron
interactions within high energies \cite{pomeranchuk}. In the
Regge theory , the  Pomeron is a colourless object having vacuum quantum
numbers. The Regge trajectory corresponds to it:
$\alpha_{\pom} (t) = \alpha_{\pom} (0) + \alpha^{\prime} t$,
where $\alpha_{\pom} (0) \simeq 1$, $\alpha^{\prime} \simeq 0.25 GeV^{-2}$
and $t$ is the $t$- channel invariant momentum transfer to the
Pomeron \cite{regge}.

In 1985, G.Ingelman and P.Schlein put forward the idea of Pomeron parton
structure \cite{ingelman_part_str}. It was then verified by the
experiments on the research of the diffractive structure function $F_2^D
(\xi, t, z, Q^2)$ \cite{ua8} - \cite{zeus_1}.

At the present time, it is supposed that $F_2^D$ can be presented as a product
of the Pomeron structure function $G_{g/ {\pom}} (z, Q^2)$ by the
factor of flow $F_ {{\pom} /p} (\xi, t)$ - the hypothesis of factorization
{\footnote {\small {There is a number of papers, in which the hypothesis of
factorization is called in question (see, for example, \cite{berera}), but
even if factorization is violated for diffractive hard scattering, the effect
can be weak at high energies \cite{kunszt}}}  }\cite{land_fact},
\cite{pritz}.  \begin{equation} F_2^D (\xi, t, z, Q^2) = F_{{\pom} /p} (\xi,
t) G_{g/ {\pom}} (z, Q^2).  \label{factor} \end{equation} The factor of flow
describes the number of Pomerons emitted by the hadron.  The Pomeron
propagator is also included in it. For proton - Pomeron interactions, the
following parametrization is most commonly used \cite{land_fact} {\footnote
{\small {Other parametrizations of the factor of flow see, for example, in
\cite{bruni}}} }:  \begin{equation} F_{{\pom} /p} (\xi, t) = \frac {N_v^2
\beta^2} {4 \pi ^2} (F_1 (t))^2 \xi^{1-2 \alpha_{\pom} (t)}.  \label{flux}
\end{equation}
Here, $\alpha_{\pom} (t)$ is the Pomeron trajectory, the
factor $N_v^2 \beta F_1 (t)$ corresponds to the proton - Pomeron interaction
vertex; $\beta = 1.8 \: GeV^{-2}$, $N_v = 3$ is the number of valence quarks
in the proton, $\xi$ is the longitudinal momentum fraction of the Pomeron,
and $F_1 (t)$ is the elastic form - factor of the proton, which is
parametrized to a high degree of accuracy by the expression:
\begin{equation} F_1 (t) = \frac {4 m_p^2 - 2.8 t} {4 m_p^2 - t} \Bigl (1 -
\frac {t} {0.7 \: GeV^2} \Bigr)^{-2},
\label{formfact}
\end{equation}
$m_p$ is the proton mass.

The choice of the Pomeron structure function $G_{g/ {\pom}} (z, Q^2)$
is determined by representing the Pomeron as a composite object.
At present, one can mark two conceptually opposite viewpoints of the
Pomeron structure {\footnote {\small {For the review see, for example,
\cite{jenk_ukraine}, \cite{kwiesinscky} and the literature in them.}} }.  According
to one of them, the Pomeron consists of $q \bar{q}$ - pairs and/or gluons,
having small fractions of momentum and slightly different in momentum from
one another. It is the so-called {\it "Soft"} or nonperturbative Pomeron.
Its intercept is equal to $\alpha_{\pom} (0) \simeq 1.085$ \cite{land_303},
\cite{land_fact}, and the evolution of the structure function is determined
by solving the equation of Dokshitzer - Gribov - Lipatov - Altarelli - Parizi
(DGLAP) \cite{dglap}. According to another point of view, the Pomeron is
composed of hard gluons and/or $q \bar{q}$ - pairs. It is the so-called
{\it "Hard"} or perturbative Pomeron. Its intercept equals $\alpha_{\pom}
(0) \simeq 1.4$ \cite{peschansky}, and the evolution of the structure
function is determined by the solution of Balitski - Fadin - Kuraev -
Lipatov's equation (BFKL) \cite{bfkl}. The available experimental data
\cite{ua8} - \cite{d0_1} do not allow for the present an unambiguous choice
to be made between them. (The mixed "DGLAP - BFKL" representation may
be actually realized. However, the integration of both equations requires
to introduce a new parameter $x_0$, ($x_0 \simeq 4 \times
10^{-3} $), at which the solutions of both equations coincide. The BFKL eq.
is solved at $x < x_0$ and the DGLAP one at $x \ge x_0$ \cite{royon}.
According to another very attractive hypothesis, there is a Pomeron
with an effective $Q^{2}$- dependence of its intercept, which leads to the
observed differences  \cite{unipom}).

Thus, a QCD - motivated study of the formation of various states in diffractive
processes at high energies has become possible due to the evolution of the primary
Pomeron hypothesis, and the following picture of such mechanisms has arisen:

i) incident hadron emits a Pomeron (this vertex is described within
the framework of the Regge theory and the factor of flow $F_{{\pom} /p} (\xi,
t)$ corresponds to it (\ref{flux}));

ii) one of the Pomeron partons is involved in hard (QCD) scattering with the
production of the investigated state.

For analytical calculations, this means that the structure function of an
interacting hadron $F_2 (x, Q^2)$ is substituted by the diffractive structure
function $F_2^D (\xi, t, z, Q^2)$ in the inclusive cross section of the
reaction with involved hadrons (\ref{factor}).

The processes of single and double diffraction are distinguished.
We are going to follow the definitions from \cite{land_303}, \cite{ingelman_part_str},
\cite{berger_qq} - \cite{schafer}. Thus, the process of hard Single Diffractive
Scattering ($SD$) is the process of hadron scattering, in which one of the
primary particles emits a Pomeron, and then it could be in principle
registered in the final state (exclusive $SD$) or not (inclusive $SD$). The
process of hard Double Diffractive Scattering ($DD$), exclusive and
inclusive, is defined in a similar way, and both initial particles emit
Pomerons.

\begin {figure} [h]
\centering
\setlength{\unitlength}{0.1in}
\psfig{figure=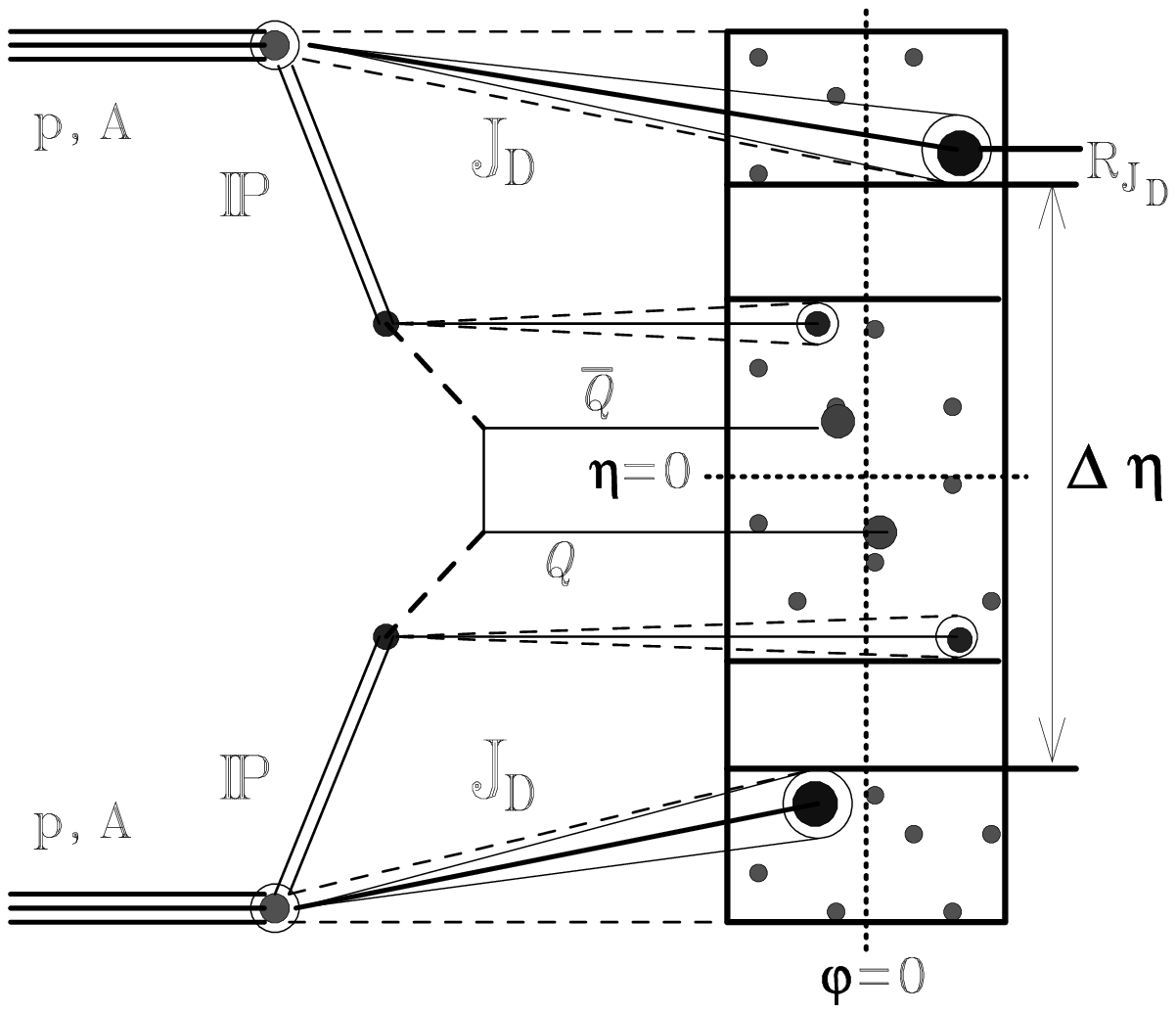,width=6.0in}
\caption {}
\label {gap}
\end {figure}

The process of uncoupled heavy quark - antiquark pair ($Q \bar{Q}$)
production in the proton - proton (DD) and Non-Coherent and Coherent ion Double
Diffractive scattering ($NDD$ and $CDD$) at high energies is considered in this
paper:
\begin{equation}
p,A \quad + \quad p,A \quad \to \quad J_{D} \quad + \quad J_{D}
\quad + \quad Q \quad + \quad \bar{Q} \quad + \quad X \quad,
\label{main_proc}
\end{equation}
where $J_{D}$ depicts a diffractive jet (see Fig. \ref{gap}).
The choice of process (\ref{main_proc}) is due to a number of reasons:

i) The cross section of this  process is larger than the production cross
section of rare states ($W$, $Z$, $H$ and so on). This facilitates its
registration, especially in the context of studying Double Diffractive
scattering and also in the research of collective nuclear
effects at high energies;

ii) It is necessary to take into account the contribution of diffractive
mechanisms for the registration of $c$ and $b$ quarks and also in consideration
of other phenomena related to the production or registration of such quarks
(for instance, the effect of heavy quarkonia suppression in QGP \cite{qgp},
the search for intermediate mass Higgs bosons \cite{higgs_gen} and so on).

\vspace{1.2cm}

{\bf \Large Diffractive production of $Q \bar{Q}$ pairs in $pp$ collisions.
Basic features}.

\vspace{0.8cm}

Let us examine the main points of the calculation of the inclusive double diffractive
$Q \bar{Q}$ production cross section for proton - proton
collisions (\ref{main_proc}) where $J_{D}$ is a diffractive jet from the
initial proton.
The production of a heavy quark pair in single diffractive interactions of
protons at high energies has explicitly been studied previously, see, for example,
\cite{land_303}, \cite{berger_qq} - \cite{duce}.

As known, the cross section of proton hard scattering (H) can be presented as:
\begin{equation}
\sigma^{H} (s_{0}) = \int_{\tau}^{1} dx_{1}
\int_{\tau / x_{1}}^{1} dx_{2} \sigma^{part.} (s) \bigl [f_{g / p_{1}}
(x_{1}, Q^{2}) f_{g / p_{2}} (x_{2}, Q^{2}) \bigr] \: ,
\label{dis}
\end{equation}
where $\sigma^{part.}$ is the cross section
of reaction (\ref{main_proc}) at a parton level \cite{sterman}, $x_{1}$ and
$x_{2}$ are the fractions of initial proton momenta carried away by partons
involved in the production of a quark pair ($x_{i} = q_{i} /k_{i}$;
$k_{i}$, $q_{i}$ are the 4-momenta of projectiles and quark (antiquark),
respectively, $i=1, 2$), $\tau = 4m_{Q}^2 / s_{0}$, and $f_{g / p_{i}}
(x_{i}, Q^{2})$ is the distribution function of parton $g$ in proton
$p_{i}$, $s_{0} = (k_{1} + k_{2})^{2}$, $M_{Q \bar{Q}} = (q_{1} +
q_{2})^{2}$.
Turning to the consideration of diffractive scattering requires to take into
account a number of additional points.

Point one. As noted above, according to the hypothesis of factorization, it is necessary to replace both partonic
distribution functions $f_{g / p_{i}} (x_{i}, Q^{2})$ by diffractive ones
(\ref{flux}) in the calculations of $DD$, i.e.:
\begin{equation} f_{g / p_{i}} (x_{i}, Q^{2}) \to
\int_{x_{i}}^{0.1} \frac {d \xi _{i}} {\xi _{i}} \int_{t_{i \:
min}}^{0} dt_{i} F_{{\pom _{i}} / p_{i}} (\xi _{i}, t_{i}) G_{g /
{\pom_{i}}} (\frac {x_{i}} {\xi _{i}}, Q^2) \: .
\label{replaced}
\end{equation}
Here $F_{{\pom _{i}} / p_{i}} (\xi _{i}, t_{i})$ is the factor of flow
(\ref{flux}), $G_{g / {\pom _{i}}} (\frac {x_{i}} {\xi _{i}}, Q^2)$ is
the Pomeron structure function, $\xi_{i}=l_{i}/k_{i}$, $l_{i}$ is the Pomeron
4-momentum, and also $s_{\pom} = (l_{1} + l_{2})^{2}$, $t_{i} = (k_{i} -
k_{i_{f}})^{2}$, where $k_{i_{f}}$ is the 4-momentum of the incident particle
after scattering (diffractive jet $J_{D_{i}}$). The evolution of the
structure functions is calculated at $Q^{2}=M_{Q \bar{Q}}^{2}$.
The upper limit to the variable $\xi_{i}$ is
determined by the condition of Pomeron exchange dominance \cite{land_fact},
and the lower limit to the variable $t_{i}$ is equal to
$t_{i \: {\it min}} = - s_{0} (1- \xi) \frac {{\it Exp} [(-1)^{i} \eta _{i}]}
{{\it Exp [\eta _{i}]} + {\it Exp} [- \eta _{i}]}$ neglecting the proton mass, where
$\eta _{i}$ is the pseudorapidity of diffractive jet $J_{D \: i}$, $i=1,2$.

As noted above, the choice of the Pomeron structure function \\
$G_{g/ {\pom}}
(z, Q^2)$ depends on that which objects, soft or hard, and also a quark
and/or a gluon, the Pomeron consists of. As the main mechanism of $Q \bar{Q}$
production at the parton level is the gluon - gluon one:
$$ g \quad + \quad g \quad \to \quad Q \quad + \quad \bar{Q} \; ,$$
gluonic distributions in the Pomeron will be of our interest. With this aim,
three parametrizations for the Pomeron gluonic structure function were chosen:
\begin{equation}
z G (z, Q_{0}^{2}) = N (1-z)^{5}, \quad \cite{ingelman_part_str}
\label{soft_g}
\end{equation}
\begin{equation}
z G (z, Q_{0}^{2}) = N z (1-z), \quad   \cite{kunszt}
\label{hard_g}
\end{equation}
\begin{equation}
z G (z, Q_{0}^{2}) = N (1-z), \quad \cite{pritz}
\label{inte_g}
\end{equation}
most commonly used in the literature {\footnote {\small {Other proposals of the
parametrization of the Pomeron structure see, for
example, in \cite{kunszt}.}} }. In all cases, the normalization constant $N$
is determined by the condition:  \begin{equation} \int_{0}^{1} zG (z,
Q_{0}^{2}) dz = 1, \label{norma} \end {equation} $Q_{0}^{2} = 4 \: GeV^2$ and
$\Lambda_{QCD} = 0.2 \: GeV$.

The second point is to take into account the conditions imposed by the
requirement of rapidity gap observation.

\begin {figure} [h]
\centering
\setlength{\unitlength}{0.1in}
\psfig{figure=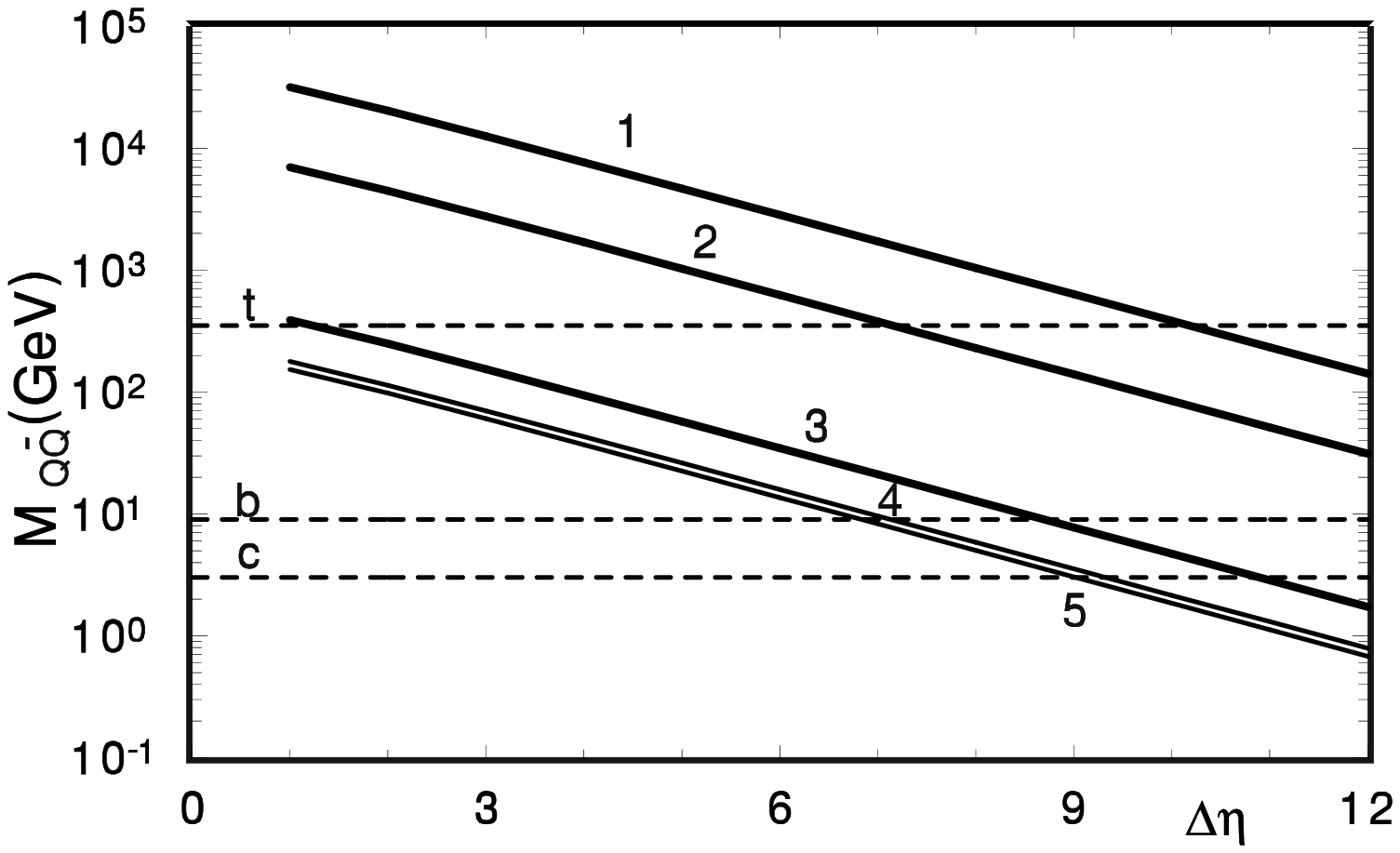,width=6.0in}
\caption {}
\label {max_mass}
\end {figure}

One can show that the pseudorapidity interval between two diffractive
jets $J_{D}$ in (\ref{main_proc}) is equal to \cite{bjerken_101},
\cite{collins}:
\begin{equation}
\Delta \eta = \eta _{2} - \eta _{1} - 2R_{J_{D}} \simeq \ln \Bigl[
\frac {s_{\pom}} {M_{Q \bar{Q}}^{2}} \Bigr] - 2R_{J_{D}} \; .
\label{gap_size}
\end{equation}
Here, $R_{J_{D}} $ is the size of the diffractive jet in the space of
azimuthal angle and pseudorapidity (assuming that $R_{J_{D}} \simeq 0.7$
\cite{cdf_1}, \cite{bjerken_101}. See Fig. \ref{gap}). In view of the
upper limit to the variable $\xi _{{\it max}} \le 0.1$, condition
(\ref{gap_size}) leads to cutting the phase volume accessible for a quark -
antiquark pair:
\begin{equation} M_{Q \bar{Q} \: {\it max}} \lsim \frac {\xi
_{{\it max}} \sqrt {s_{0}}} {{\it Exp} (\Delta \eta / 2 + R_{J_{D}})}.
\label{mass_dep}
\end{equation}
Fig. \ref{max_mass} presents the $M_{Q \bar{Q} \: max}$ dependence on
$\Delta \eta$ for different values of $\sqrt {s_{0}}$. As an example, one can
see from the figure that the fulfilment of condition (\ref{mass_dep}) results
in that the double diffractive production of a pair of $t$ quarks is prohibited
at Tevatron, and their observation in usual $DD$ at LHC is hardly probable.
A similar result was obtained in \cite{heyssler_qq}. At the same time, in
collective interactions of ions at LHC, condition (\ref{mass_dep}) can be
fulfilled for  $t \bar{t}$, $WW$, $ZZ$ pairs, and such states can be observed
\cite{my_zz}.

In the above reasonings, it is supposed that the pseudorapidity intervals,
corresponding to Pomeron exchange (see. Fig. \ref{gap}), remain empty.
The situation is actually such that these sites turn out to be
filled with a certain number of secondary hadrons. This might be due to
statistical fluctuations: "leakage through the gap edges" (i.e., there exists
a certain probability that hadrons, expected close to the direction of incident
particles, occur in the central area fluctuatively) or the result of interaction
of other, not "Pomeronic", partons of initial particles - "multiple
interaction". These effects are of particular importance when nucleus -
nucleus interactions are considered. To take them into account, it was
proposed to introduce an additional factor into the expression for diffractive
cross section - "survival probability" $<|S|^{2}>$ \cite{bjerken_101}
defined as follows. If $|S(s,b)|^{2}$ is the probability that
two hadrons pass one through another with impact parameter $b$ at given
$\sqrt{s}$ without interactions, except a hard one, the survival probability
can be written as:
\begin{equation} <|S|^{2}> = \frac {\int F(b)
|S(s,b)|^{2} d^{2} b} {\int F(b) d^{2} b} \;,
\label{survey}
\end{equation}
where $F(b)$ is the usual overlap of partonic densities of interacting hadrons
in the space of the impact parameter. The evaluations of $<|S|^{2}>$ in various
approximations: eikonal, gaussian and some others, give the value of the order
of $\sim 5 \div 23 \%$ at LHC energy \cite{bjerken_101}, \cite{survival}.
With increasing $\sqrt{s}$, the value of $<|S|^{2}>$ decreases, i.e. it becomes
less probable that two hadrons will not interact at high energies (the total
cross  section $\sigma _{\it {tot}}$ is proportional to the region of soft
interaction $\pi R^{2}$ in which diffractive scattering happens; this region
is in its turn proportional to $ln \: s$).

Following the currently accepted estimates, $<|S|^{2}> \simeq 10\%$ at
$\sqrt{s} = 14 TeV$ for proton - proton interactions.

It should be also emphasized that the filling of rapidity gaps can be due to
the so-called "{\it pile up}" - effect (the effect of "superposition") when
some proton - proton interactions occur in a very small volume.
It is obvious that it depends on the intensities of
interacting beams.  This effect should be kept in mind for LHC. The
degree of its influence on experimental results depends on the properties of
a registering device. (This problem will be considered in future.)

\begin {figure} [h]
\centering
\setlength{\unitlength}{0.1in}
\psfig{figure=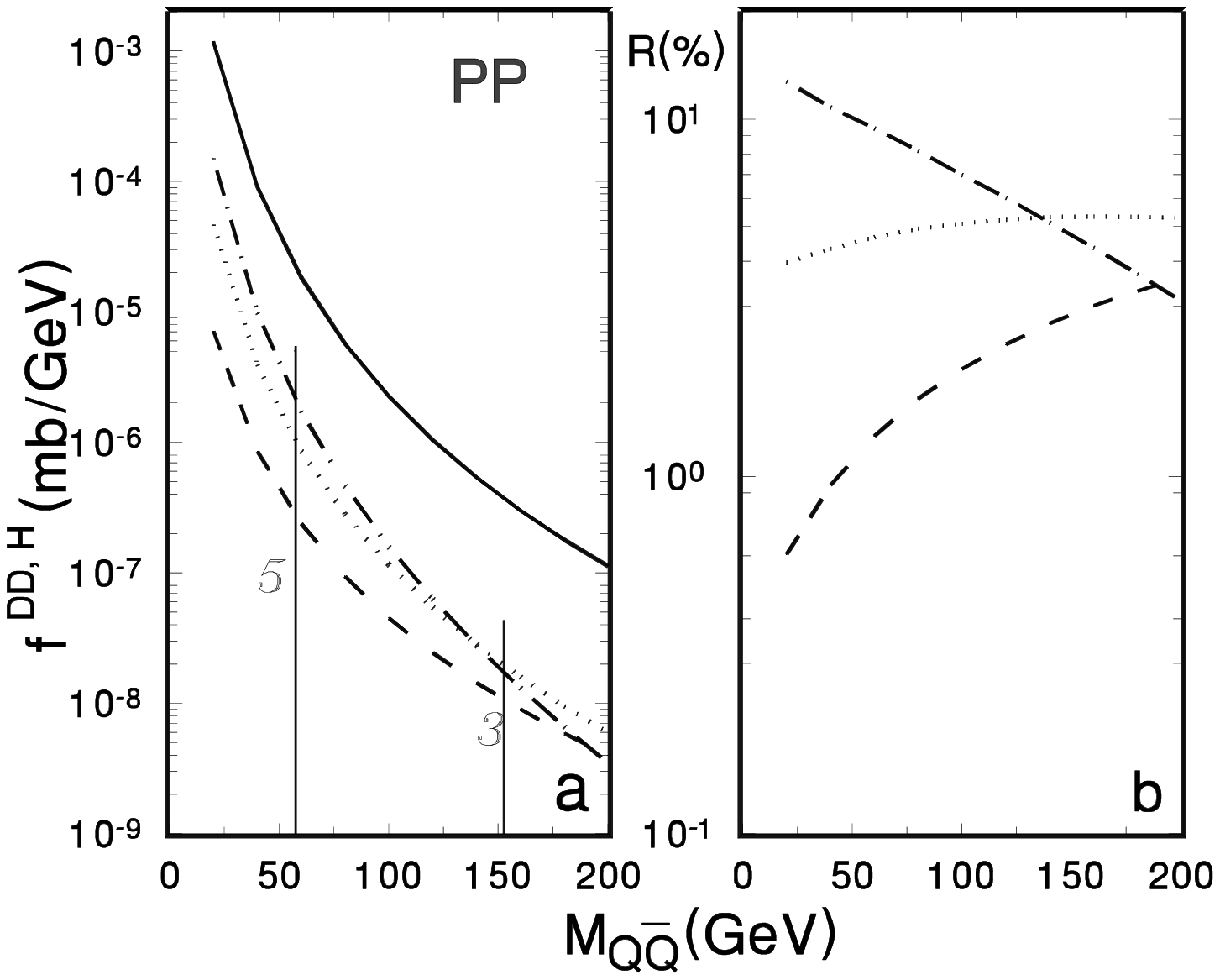,width=6.0in}
\caption {}
\label {pp_fig}
\end {figure}

Fig. \ref{pp_fig} presents the calculated results of the differential cross
section for quark - antiquark pair production in hard ($H$) and Double
Diffractive ($DD$) scatterings of protons:
\begin{equation} f^{DD,H} = \frac {d^{3}\sigma^{DD,H}} {dM_{Q \bar{Q}}
d\eta_{Q} d\eta_{\bar{Q}}} \;,
\label{diff_cross}
\end{equation}
and the ratio
\begin{equation} R = \frac {d^{3}\sigma^{DD}} {dM_{Q \bar{Q}}
d\eta_{Q} d\eta_{\bar{Q}}}  / \frac {d^{3}\sigma^{H}} {dM_{Q \bar{Q}}
d\eta_{Q} d\eta_{\bar{Q}}} 100 \% \;,
\label{r}
\end{equation}
versus the invariant mass of quark pair $M_{Q \bar{Q}}$ at $\eta_{Q} =
\eta_{\bar{Q}} =0$ and $\sqrt{s_{0}} =14 TeV$ for the chosen models of
the Pomeron structure function (\ref{soft_g}) - (\ref{inte_g}), where
$\eta_{Q,\bar{Q}}$ are the pseudorapidities of heavy quarks $Q, \bar{Q}$. The
solid line corresponds to the evaluation made for hard (QCD) scattering and
the dot - dashed, dashed and dotted lines correspond to the evaluations of
$DD$ made using Pomeron models (\ref{soft_g}), (\ref{hard_g}) and
(\ref{inte_g}), respectively. The upper limits on $M_{Q \bar{Q}}$ for different
values of rapidity gap $\Delta \eta$, calculated from (\ref{mass_dep}), are
denoted by straight lines. The factor "survival probability" (\ref{survey}) is
not taken into account {\footnote {\small {The factor (\ref{survey}) does not
affect the number of produced quark pairs and it must be taken into account
when rapidity gaps are only extracted. }} }.

As seen from the figure, the behaviour of the differential cross
sections is noticeably different for the considered models of Pomeron. It
promotes their recognition in an experiment. The difference in the
cross sections is greater than one order of magnitude for the "{\it Soft}"
(\ref{soft_g}) and "{\it Hard}" (\ref{hard_g}) models in the region of small
invariant masses ($M_{Q \bar{Q}} \lsim 50 GeV$) and it decreases equalizing
at $M_{Q \bar{Q}} \sim 140 \div 180 GeV$ for all the models.

As seen from the figure, the production of quark pairs with a large
invariant mass and  a large gap between diffractive jets
(for example, $M_{Q \bar{Q}} \gsim 50 GeV$ and $\Delta \eta > 5$) is
forbidden.  Remind that the maximum allowable value of $M_{Q \bar{Q}}$
for Double Diffractive $Q \bar{Q}$ production at $\sqrt{s_{0}} =14 TeV$
is approximately $500 GeV$ as it follows from (\ref{mass_dep}).

The fraction of diffractive pairs in the considered kinematical region is an
average of $2 \div 8 \%$  of the number of those produced in the hard (QCD)
process {\footnote {\small {It is obvious that this number will be
greater for Single Diffraction.}} }. This is enough for studying diffractive
physics with their aid. At the same time, this fraction is very large and
diffractive mechanisms should be taken into account to detect
heavy quarks at least in the region of $M_{Q \bar{Q}} \lsim 200 \div 500 GeV$.

The total cross sections of heavy quark pair production are obtained by
integration (\ref{dis}) over all the variables taking (\ref{replaced}) into
account. The results are presented in Table \ref{pp_tab}. It is seen that the
fraction of heavy quarks, produced in $DD$, is $\sim 1 \div 18 \%$ of the
number of those produced in hard scattering. The results
obtained in \cite{heyssler_qq} without taking into account the factor
$<|S|^{2}>$ for $\sqrt{s_{0}} =10 TeV$ are presented for comparision.
The data obtained by us are a little bit higher than the ones from \cite{heyssler_qq}.
This rather well explained by choosing another proton structure function \cite{owens}.

\vspace{1.2cm}

{\bf \Large Double diffractive production of $Q \bar{Q}$ pairs in
collisions of heavy ions.}

\vspace{0.8cm}

Let us consider the process of quark - antiquark pair production in the
double diffractive scattering of heavy ions. In this case, the calculations
are carried out similar to the above described $DD$ of protons, taking into
account that the proton form - factor (\ref{formfact})  for coherent scattering
is substituted by the nucleus form - factor \cite{schafer} which we parametrize
as \cite{kaidalov}:
\begin{equation}
F(t) \sim exp (\: R_{A}^{2} \: t \: / \: 6 \:) \:,
\label{ion_formfac}
\end{equation}
where $R_{A}$ is the radius of the nucleus $A$ ($R_{A} = r_{0} A^{1/3}$,
$r_{0} = 1.2 fm$).

The research of the $A$ - dependence of the non-coherent diffractive scattering
of symmetric nuclei allows the
cross section to be parametrized as \cite{kaidalov}, \cite{capella}:
\begin{equation}
\sigma_{A} \simeq A^{2 \alpha} \sigma_{N} \;,
\label{parametrization}
\end{equation}
where the exponent $\alpha \sim 0.7 \div 0.8$ for peripheral (diffractive)
processes \cite{capella} and $\alpha \sim 0.95 \div 1.0$ for central (hard)
ones, $\sigma_{N}$ is the nucleon - nucleon cross section. We used
$\alpha = 0.7$ for $NDD$ and $\alpha = 1.0$ for hard ion scattering.

\begin {figure} [h]
\centering
\setlength{\unitlength}{0.1in}
\psfig{figure=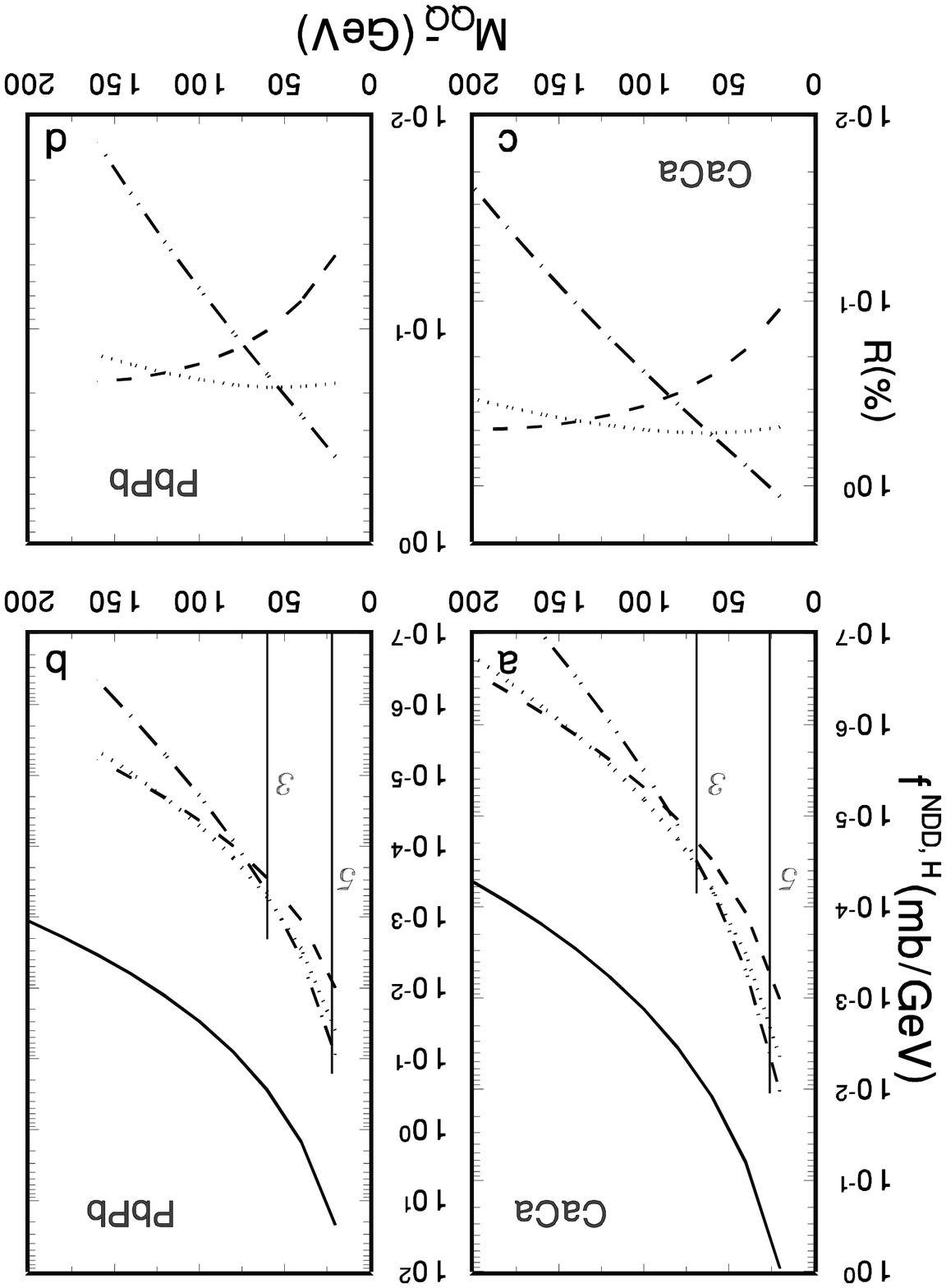,width=6.0in}
\caption {}
\label {ion_fig_non}
\end {figure}

Fig. \ref{ion_fig_non} presents the dependence of the differential cross
section $f^{NDD,H}$ (\ref{diff_cross}) and ratio $R(\%)$ (\ref{r}) on the
invariant mass $M_{Q \bar{Q}}$ for quark - antiquark pair production in the
Hard ($H$) and Non-coherent Double Diffractive ($NDD$) scattering of
$Ca$ and $Pb$ ions for the chosen models of Pomeron (\ref{soft_g}) - (\ref{inte_g}).
It should be noted that the factor (\ref{survey}) was not
taken into account as in case of proton interactions.
Note that there are large differences in estimating the value of $<|S|^{2}>$
(\ref{survey}) with a variety of approaches. (This problem will be considered in
future.) It should be emphasized that the value of $<|S|^{2}>$ will be much
smaller than for proton - proton collisions in case of $NDD$ because of
multiple collisions of projectile nucleons.

As seen from the figure, the behaviour of the $NDD$ differential cross section for
all models of the Pomeron stucture function is similar to the ones in case of
proton interactions at $\sqrt{s_{0}}= 14 TeV$. However, the difference between
the models is a little bit smaller because of small values of $\sqrt{s_{0}}$
($\lsim 10$ at $M_{Q \bar{Q}} \sim 20 GeV$). This leads to that the models
become indistinguishable at $M_{Q \bar{Q}} \sim 50 \div 100 GeV$. Then, the
difference rises rapidly due to that the {\it "Soft"} Pomeron (\ref{soft_g}) is characterized
by a sharper fall of the cross section than the {\it "Hard"} one (\ref{hard_g})
with increasing the invariant mass and reaches one order of magnitude
at $M_{Q \bar{Q}} \sim 160 \div 200 GeV$. As seen from the figure, the
production of quark pairs with a large invariant mass and large
rapidity gaps (for instance, $M_{Q \bar{Q}} \gsim 30 GeV$ and
$\Delta \eta \ge 5$) is forbidden.

Table \ref{ion_tab_non} shows the total cross sections of the hard (central) and
non-coherent double diffractive production of heavy quarks and their ratios to
the hard one for selected models (\ref{soft_g}) - (\ref{inte_g}) in $CaCa$ and
$PbPb$ interactions. As seen from the table, the fraction of diffractive
$Q \bar{Q}$ pairs averages approximately $\sim 1(0.01) \%$ of the number of
such pairs produced in central $Ca(Pb)$ collisions.

The coherent double diffractive scattering ($CDD$) was calculated by
replacing the proton form - factor (\ref{formfact}) by the form -
factor of the nucleus \cite{kaidalov}. Thus, the total energy of the interacting
system is
\begin{equation}
\sqrt{s_{A}} = A \sqrt{s_{0}} \:,
\label{ros_ion}
\end{equation}
neglecting the nucleus mass,
where $\sqrt{s_{0}}$ is the total energy of nucleon - nucleon interactions
($\sqrt{s_{0}}=5.5 TeV$ for $PbPb$ beams and $\sqrt{s_{0}}=6.3 TeV$ for $CaCa$
ones \cite{lhc_desine}).

\begin {figure} [h]
\centering
\setlength{\unitlength}{0.1in}
\psfig{figure=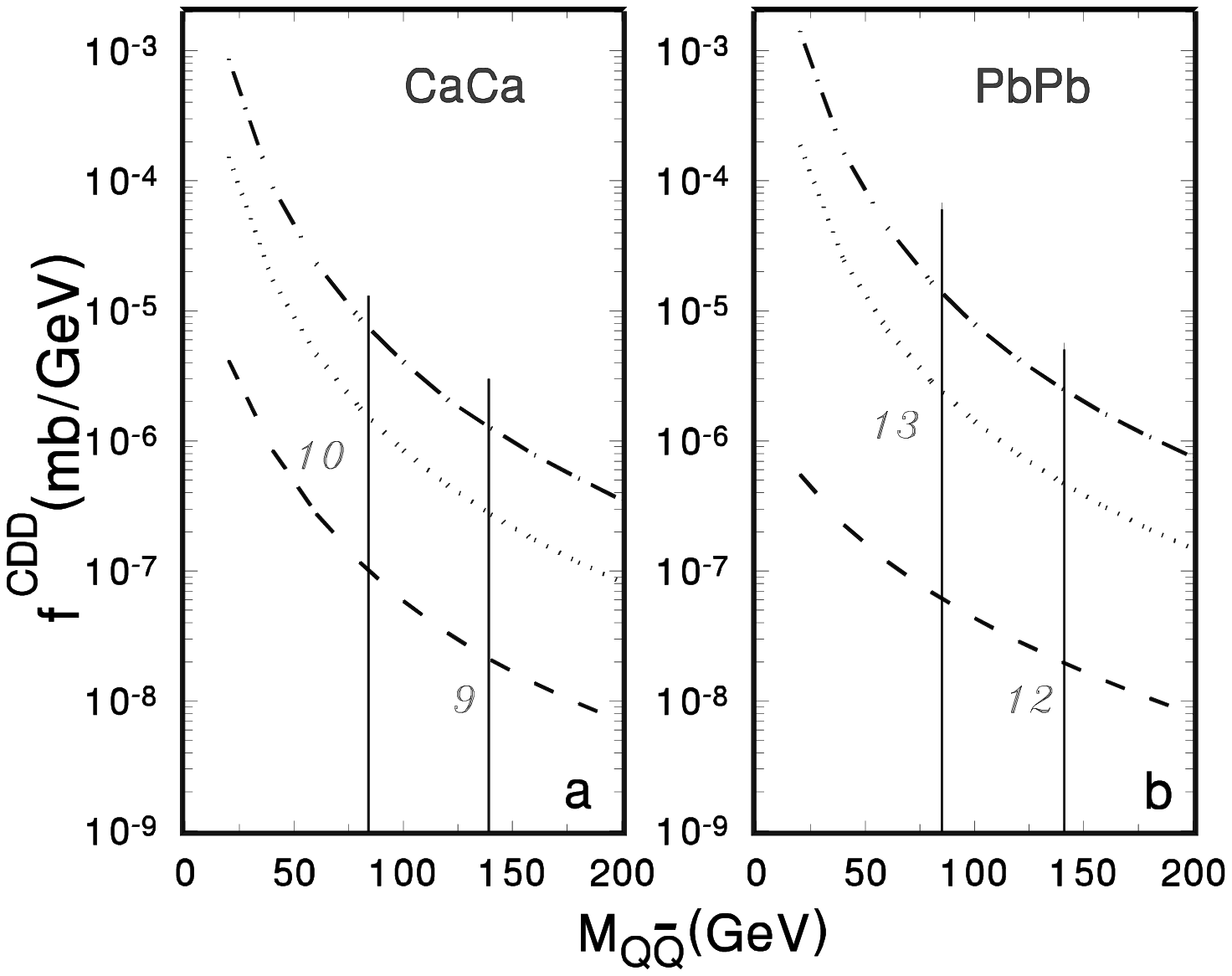,width=6.0in}
\caption {}
\label {ion_fig_coh}
\end {figure}

Fig. \ref{ion_fig_coh} depicts the differential cross section $f^{CDD}$
(\ref{diff_cross}) versus mass $M_{Q \bar{Q}}$ in the coherent double
diffractive scattering of $Ca$ and $Pb$ ions for the selected Pomeron models.
In this case, the factor
(\ref{survey}) was not considered in the calculations of the total and
differential cross sections. (It sould be stressed that the value of
$<|S(s,b)|^{2}>$ differs from the case of non-coherent scattering here.)
As in the previous cases, the upper limits on $M_{Q \bar{Q}}$ at different
$\Delta \eta$ are denoted by straight lines.

As seen from the figure, in contrast to the previous cases, there is a smoother
fall of the cross section with increasing $M_{Q \bar{Q}}$. Thus, if the
non-coherent  cross section of $Q\bar{Q}$ production at small $M_{Q \bar{Q}}$
($M_{Q \bar{Q}} \lsim 50 GeV$) is larger than the coherent one by one - four
orders of magnitude, these cross sections become equal at higher invariant
masses ($M_{Q \bar{Q}} \sim 100 \div 160 GeV$), and  the coherent cross section becomes larger than the non-coherent
one for $Ca$ and $Pb$ beams at still higher values of $M_{Q \bar{Q}}$, for the
considered Pomeron models. This difference is stronger for the model of $"Soft"$
Pomeron (\ref{soft_g}). Such a behaviour follows from a threshold fall of the
non-coherent cross section at large $M_{Q \bar{Q}}$. Including the total energy
of the nucleus (\ref{ros_ion}) in the interaction shifts this threshold to the
area of very large invariant masses at given size of a rapidity gap (see Fig.
\ref{max_mass}).
Thus, one can formulate the observation conditions of collective nuclear
interactions: the detection of a heavy quark - antiquark pair with a
large invariant mass and a large rapidity gap between diffractive jets (for
example, $M_{Q \bar{Q}} > 100 GeV$ at $ \Delta \eta >3$ or $M_{Q \bar{Q}} >
50 GeV$ at $ \Delta \eta >5$ etc.) The coherent cross
section is $\sim 10^{-5} \div 10^{-7}mb$ in these regions. At the luminosities
of heavy ions planned at LHC ($\sim 10^{26}$ for $PbPb$ and $\sim 10^{30}$ for
$CaCa$ \cite{eggert}), this  allows one to have up to $10^{4}$ such events
in a 15- day run ($10^{6}s$) of the collider. Medium nuclei ($A < 100$) are
more preferable because they have higher luminosities, and so they give
larger number of events. Note that the
multiplicities of secondary particles are lower in the collisions of
light and medium nuclei. It would promote the extraction of a rapidity gap. At
the same time, light nuclei do not allow one to move far off the threshold
of non-coherent production what complicates the isolation of a pure
coherent contribution to the diffractive cross section.

As seen from the figure, the cross sections of coherent scattering differ
markedly for the Pomeron models under consideration. Thus, this difference is
$2 \div 3$ orders of magnitude for "{\it Soft}" (\ref{soft_g}) and "{\it Hard}"
(\ref{hard_g}) Pomerons and a little bit smaller than one order for the models
(\ref{soft_g}) and (\ref{inte_g}). Such a behaviour of the cross sections
changes weakly in the considered region of invariant masses ($20GeV \le
M_{Q \bar{Q}} \le 200 GeV$). As in the previous cases, the cross section of
"{\it Hard}" Pomerons falls smoother than that for "{\it Soft}" ones. However,
the region, where the differential cross sections of the studied models
coincide, lies far to the higher $M_{Q \bar{Q}}$ in distinction to proton -
proton and non-coherent ion interactions.

The total cross sections of $c \bar{c}$ and $b \bar{b}$ pair production in the
coherent double diffractive scattering of $CaCa$ and $PbPb$ are given in Table
\ref{ion_tab_coh}. It sould be noted that the total cross section of the
"{\it Hard}" model (\ref{hard_g}) is larger for $CaCa$ interactions than for
$PbPb$ ones.  The situation is opposite for the models (\ref{soft_g}) and
(\ref{inte_g}).

\vspace{1.2cm}

{\bf \Large Conclusions.}

\vspace{0.8cm}

As it follows from the foregoing, the number of heavy quark pairs produced in
double (and, moreover, single) diffractive scattering is a major part of those
produced in hard (central) interactions at LHC energies. Therefore,  the
contribution of diffractive mechanisms should be taken into account as an
additional process for the registration of heavy quarks and in the studies of
phenomena related to their production or registration (for instance,
heavy quarkonia suppression in QGP, search for intermediate mass Higgs boson and
so on). At present, the contribution of diffractive mechanisms is small because
the values of $\sqrt{s_{0}}$ of the running colliders are small. Therefore, a
disagreement between theoretical estimates and experimental results is
insignificant. It is hoped that the diffractive processes will play an important
role in looking for rare states (Higgs particles, new gauge bosons, heavy quarkonia, etc)
due to their distinctive  features.

On the other hand, the cross sections of diffractive particle production
at LHC energies are large enough, and so one can study different aspects of
diffractive physics at high energies, namely, the behaviour of
structure functions at small $x$, the hypothesis of factorization, nuclear
shadowing and some other collective nuclear phenomena. Such investigations can
be made using the CMS or FELIX setups. They will cover a large interval
of pseudorapidity ($\sim 10$ and $> 14$ units, respectively) and will have
detectors with a high resolution in the central region. Thus, the diffractive
cones will be covered and the centrally produced states will be detected, i.e. the
complete event could be reconstructed.

The considered invariant mass interval of a centrally produced
$Q \bar{Q}$ pair ($20 GeV \le M_{Q \bar{Q}} \le 200 GeV$) is apparently the
most optimum one for studying double diffraction at LHC. The cross sections,
as well as the masses of produced states, are large enough, which makes for
certain event registration.  Furthermore, the areas in which the
cross sections of the considered models of Pomeron differ significantly for
all types of interactions, lie in the mentioned interval of $M_{Q
\bar{Q}}$.

Finally, note that the study of diffractive interaction is particularly urgent
at the first stage of the collider operation, when the beam focusing and
luminosity have not reached their designed values yet (i.e., when the fraction
of diffractive interactions is larger) and when light and medium nuclei are
accelerated.

The authors express their sincere gratitude to M.G.Hayrapetyan and
E.A.Strokovsky for fruitful discussions and valuable remarks.



\newpage

{\bf Table Captions.}

\bigskip

Table \ref{pp_tab}.
Total cross sections (mb) of $c \bar{c}$ and $b \bar{b}$
pair production in the processes of Hard (QCD) ($\sigma^{H}$) and Double
Diffractive ($\sigma^{DD}$) scattering of protons at different $\sqrt{s_{0}}$
using models (\ref{soft_g}), (\ref{hard_g}) and (\ref{inte_g}) of the Pomeron
structure function. The gluon distribution in the proton
is taken from \cite{owens}. The ratio $R(\%)$ of the total diffractive cross
section to the hard one for each model is also given. The results of
\cite{heyssler_qq} obtained at $\sqrt{s_{0}} = 10 TeV$ for model (\ref{hard_g})
is presented for comparison. The factor (\ref{survey}) is not taken into
account.

\vspace{0.5cm}

Table \ref{ion_tab_non}.
Total cross sections (mb) of $c \bar{c}$ and $b \bar{b}$
pair production in the processes of central $\sigma^{H}_{A}$ and Non-Coherent
Double Diffractive $\sigma^{NDD}$ scattering of $Ca$ and $Pb$ ions using
the Pomeron models (\ref{soft_g}), (\ref{hard_g}) and (\ref{inte_g}). The
ratio $R(\%)$ of the total diffractive cross section to the central one
for each model is given too. The factor (\ref{survey}) is not taken into
account.

\vspace{0.5cm}

Table \ref{ion_tab_coh}.
Total cross sections $\sigma^{CDD}$(mb) of $c \bar{c}$ and $b \bar{b}$ pair
production in the Coherent Double Diffractive scattering of
$Ca$ and $Pb$ ions using the above models of Pomeron. The factor
(\ref{survey}) is not taken into account.

\newpage

{\bf Figure Captions.}

\bigskip

Figure \ref{gap}.
Schematic representation of heavy quark - antiquark ($Q \bar{Q}$) production
in Double Diffractive scattering (\ref{main_proc}) using the hypothesis of
factorization in the plane of azimuthal angle ($\phi$) and pseudorapidity
($\eta$). Projectiles (protons - $p$ or nuclei - $A$) emit Pomerons ($\Pom$)
and escape producing the diffractive jets ($J_{D}$), the interval between
which ($\Delta \eta$) is a "rapidity gap".  The quark - antiquark pair ($Q
\bar{Q}$) is produced from the hard (QCD) interaction of both Pomerons. The
radius of the diffractive cones in ($\eta,\phi$) is taken equal to $R_{J_{D}}
\simeq 0.7$.

\bigskip

Figure \ref{max_mass}.
Upper (solid lines) and lower (dashed lines) limits on the invariant mass of
double diffractively produced heavy quark - antiquark pair $M_{Q \bar{Q}}(GeV)$
versus rapidity gap size $\Delta \eta$ for different types of interactions and
produced particles: 1 -$PbPb$ Coherent ($\sqrt{s_{0}}=1144TeV$), 2 -$CaCa$
Coherent ($\sqrt{s_{0}}=252TeV$), 3 -$pp$ ($\sqrt{s_{0}}=14TeV$), 4 -$CaCa$
Non-Coherent ($\sqrt{s_{0}}=6.3TeV$ per nucleon), 4 -$PbPb$ Non-Coherent
($\sqrt{s_{0}}=5.5TeV$ per nucleon), t - $t \bar{t}$ ($m_{t}= 175 GeV$),
b - $b \bar{b}$ ($m_{b}= 4.5 GeV$), c- $c \bar{c}$ ($m_{c}= 1.5 GeV$).

\bigskip

Figure \ref{pp_fig}.
Dependence of {\bf (a)} the differential cross section of quark - antiquark pair
production $f^{DD,H} (mb/GeV) = \frac {d^{3} \sigma^{DD,H}}{dM_{Q \bar{Q}}
d \eta_{Q} d \eta_{\bar{Q}}}$ at $\eta_{Q} = \eta_{\bar{Q}}=0$ (\ref{diff_cross})
and {\bf (b)} the ratio $R(\%)=f^{DD}/f^{H}$ (\ref{r}) on the invariant mass of
quark pair $M_{Q \bar{Q}}(GeV)$ for the chosen models of Pomeron: (\ref{soft_g}) -
dot - dashed lines, (\ref{hard_g}) - dashed lines, (\ref{inte_g}) - dotted lines,
solid  line is for hard (QCD) proceess. $\sqrt{s_{0}} =14 TeV$, the factor
(\ref{survey}) is not taken into account. The upper limits on $M_{Q \bar{Q}}$
for $\Delta \eta =$ 3 and 5 are denoted by vertical lines.

\bigskip

Figure \ref{ion_fig_non}.
The same as in Fig. \ref{pp_fig} only for non-coherent scattering of $CaCa$
at $\sqrt{s_{0}}= 6.3 TeV$ {\it per nucleon} {\bf (a,c)} and $PbPb$ at
$\sqrt{s_{0}}= 5.5 TeV$ {\it per nucleon} {\bf (b,d)}. The factor
(\ref{survey}) is not taken into account.

\bigskip

Figure \ref{ion_fig_coh}.
Dependence of the differential cross section $f^{CDD}(mb/GeV)$ (\ref{diff_cross})
of $Q \bar{Q}$ production in the coherent double diffractive scattering of $CaCa$
{\bf (a)} and $PbPb$ {\bf (b)} on the invariant mass of pair $M_{Q \bar{Q}}(GeV)$
at $\eta_{Q} = \eta_{\bar{Q}} =0$ for the chosen models of Pomeron. The
designations are the same as in the previous figures. The factor
(\ref{survey}) is not taken into account. The upper limits on $M_{Q \bar{Q}}$
are denoted for $\Delta \eta=$ 9,10 ($CaCa$) and $\Delta \eta=$ 12,13
($PbPb$) by straight lines.

\newpage

\begin{table}[t]
\begin{center}
\begin{tabular}{|c|c|c|}
\hline
$\sqrt{s_{0}}$ = 14 TeV              & $   c \bar{c}          $ & $    b \bar{b} $  \\
\hline
$\sigma^{H}$                         & $ 1.63 \times 10^{-2}  $ & $ 4.49 \times 10^{-3} $ \\
\hline
$\sigma^{DD}$, model (\ref{soft_g}) & $ 2.90 \times 10^{-3}  $ & $ 2.48 \times 10^{-4} $ \\
\hline
$ R (\%) = \sigma^{DD} / \sigma^{H}$ & $   17.79 $ & $   5.52          $ \\
\hline
$\sigma^{DD}$, model (\ref{inte_g}) & $ 1.40 \times 10^{-3}  $ & $ 1.605 \times 10^{-4} $ \\
\hline
$ R (\%) = \sigma^{DD} / \sigma^{H}$ & $   8.59          $ & $   3.575         $ \\
\hline
$\sigma^{DD}$, model (\ref{hard_g}) & $ 1.26 \times 10^{-4}  $ & $ 2.76 \times 10^{-5} $ \\
\hline
$ R (\%) = \sigma^{DD} / \sigma^{H}$ & $   0.77          $ & $   0.615         $ \\
\hline
\hline
$\sqrt{s_{0}}$ = 10 TeV \cite{heyssler_qq} &               & \\
\hline
$\sigma^{H}$                               & $ 9.65 \times 10^{-3} $ & $ 2.91
\times 10^{-3} $ \\ \hline $\sigma^{DD}$, model (\ref{hard_g}) & $ 6.56
\times 10^{-5}  $ & $ 1.51 \times 10^{-5} $ \\
\hline
$ R (\%) = \sigma^{DD} / \sigma^{H}$       & $   0.68          $ & $   0.52          $ \\ \hline
\hline
$\sqrt{s_{0}}$ = 6.3 TeV                   &               & \\
\hline
$\sigma^{H}$                               & $ 5.12 \times 10^{-3} $ & $ 1.65 \times 10^{-3} $ \\
\hline
$\sigma^{DD}$, model (\ref{soft_g}) & $ 3.90 \times 10^{-4}  $ & $ 2.75 \times 10^{-5} $ \\
\hline
$ R (\%) = \sigma^{DD} / \sigma^{H}$       & $   7.62 $ & $   1.67          $ \\
\hline
$\sigma^{DD}$, model (\ref{inte_g})       & $ 2.30 \times 10^{-4}  $ & $ 2.545 \times 10^{-5} $ \\
\hline
$ R (\%) = \sigma^{DD} / \sigma^{H}$       & $   4.49          $ & $   1.54          $ \\
\hline
$\sigma^{DD}$, model (\ref{hard_g})       & $ 3.27 \times 10^{-5}  $ & $ 7.68 \times 10^{-6} $ \\
\hline
$ R (\%) = \sigma^{DD} / \sigma^{H}$       & $   0.64          $ & $   0.465         $ \\
\hline
\hline
$\sqrt{s_{0}}$ = 5.5 TeV                   &                     & \\
\hline
$\sigma^{H}$                               & $ 4.27 \times 10^{-3}  $ & $ 1.39 \times 10^{-3} $ \\
\hline
$\sigma^{DD}$, model (\ref{soft_g})       & $ 2.85 \times 10^{-4}  $ & $ 1.90 \times 10^{-5} $ \\
\hline
$ R (\%) = \sigma^{DD} / \sigma^{H}$       & $   6.675 $ & $   1.37          $ \\
\hline
$\sigma^{DD}$, model (\ref{inte_g})       & $ 1.75 \times 10^{-4}  $ & $ 1.895 \times 10^{-5} $ \\
\hline
$ R (\%) = \sigma^{DD} / \sigma^{H}$       & $   4.10 $ & $   1.36          $ \\
\hline
$\sigma^{DD}$, model (\ref{hard_g})       & $ 2.71 \times 10^{-5}  $ & $ 6.38 \times 10^{-6} $ \\
\hline
$ R (\%) = \sigma^{DD} / \sigma^{H}$       & $   0.635         $ & $   0.46          $ \\
\hline
\end{tabular}
\end{center}
\caption{  }
\label{pp_tab}
\end{table}

\newpage

\begin{table}[t]
\begin{center}
\begin{tabular}{|c|c|c|}
\hline
$CaCa$,   Non-Coherent                   & $   c \bar{c} $ & $ b \bar{b} $  \\
\hline
$\sigma^{H}_{A}$                          & $ 8.192                $ & $ 2.64                $ \\
\hline
$\sigma^{NDD}$, model (\ref{soft_g})     & $ 6.82 \times 10^{-2}  $ & $ 4.81 \times 10^{-3} $ \\
\hline
$ R (\%) = \sigma^{NDD} / \sigma^{H}_{A}$ & $   0.83 $ & $   0.18          $ \\
\hline
$\sigma^{NDD}$, model (\ref{inte_g})     & $ 4.02 \times 10^{-2}  $ & $ 4.45 \times 10^{-3} $ \\
\hline
$ R (\%) = \sigma^{NDD} / \sigma^{H}_{A}$ & $   0.49          $ & $   0.17          $ \\
\hline
$\sigma^{NDD}$, model (\ref{hard_g})     & $ 5.72 \times 10^{-3}  $ & $ 1.34 \times 10^{-3} $ \\
\hline
$ R (\%) = \sigma^{NDD} / \sigma^{H}_{A}$ & $   0.07          $ & $   0.05         $ \\
\hline
\hline
$PbPb$,   Non-Coherent                   &                     &         \\
\hline
$\sigma^{H}_{A}$                          & $ 184.74           $ & $ 60.14          $ \\
\hline
$\sigma^{NDD}$, model (\ref{soft_g})     & $ 5.01 \times 10^{-1}  $ & $ 3.34 \times 10^{-2} $ \\
\hline
$ R (\%) = \sigma^{NDD} / \sigma^{H}_{A}$ & $   0.27  $ & $   0.06          $ \\
\hline
$\sigma^{NDD}$, model (\ref{inte_g})     & $ 3.08 \times 10^{-1}  $ & $ 3.33 \times 10^{-2} $ \\
\hline
$ R (\%) = \sigma^{NDD} / \sigma^{H}_{A}$ & $   0.17  $ & $   0.05          $ \\
\hline
$\sigma^{NDD}$, model (\ref{hard_g})     & $ 4.77 \times 10^{-2}  $ & $ 1.12 \times 10^{-2} $ \\
\hline
$ R (\%) = \sigma^{NDD} / \sigma^{H}_{A}$ & $   0.026         $ & $   0.019         $ \\
\hline
\end{tabular}
\end{center}
\caption{  }
\label{ion_tab_non}
\end{table}

\vspace{2cm}

\begin{table}[t]
\begin{center}
\begin{tabular}{|c|c|c|}
\hline
$CaCa$,  Coherent & $   c \bar{c} $ & $ b \bar{b} $  \\
\hline
$\sigma^{CDD}$, model (\ref{soft_g}) & $ 3.58 \times 10^{-2}  $ & $ 7.27 \times 10^{-4} $ \\
\hline
$\sigma^{CDD}$, model (\ref{inte_g}) & $ 1.14 \times 10^{-2}  $ & $ 2.56 \times 10^{-4} $ \\
\hline
$\sigma^{CDD}$, model (\ref{hard_g}) & $ 1.01 \times 10^{-4}  $ & $ 6.15 \times 10^{-6} $ \\
\hline
\hline
$PbPb$,  Coherent &              &                   \\
\hline
$\sigma^{CDD}$, model (\ref{soft_g}) & $ 6.01 \times 10^{-2}  $ & $ 1.43 \times 10^{-3} $ \\
\hline
$\sigma^{CDD}$, model (\ref{inte_g}) & $ 1.72 \times 10^{-2}  $ & $ 4.43 \times 10^{-4} $ \\
\hline
$\sigma^{CDD}$, model (\ref{hard_g}) & $ 1.96 \times 10^{-5}  $ & $ 2.46 \times 10^{-6} $ \\
\hline
\end{tabular}
\end{center}
\caption{ }
\label{ion_tab_coh}
\end{table}

\newpage

\newpage

\newpage

\newpage

\newpage

\end {document}